\documentclass{moriond}

\usepackage{amssymb}

\def\be{\begin{equation}}
\def\ee{\end{equation}}
\def\bea{\begin{eqnarray}}
\def\eea{\end{eqnarray}}

\newcommand{\Photo}{}

\begin{document}
\vspace*{4cm}
\title{Impact of Observational and Modelling Assumptions on Intergalactic Magnetic Field Constraints from TeV Gamma-Ray Bursts with the Cherenkov Telescope Array Observatory}

\author{Ténéman Keita, Renaud Belmont, Thierry Stolarczyk}

\address{Université Paris-Saclay, Université Paris Cité, CEA, CNRS, AIM, 91191, Gif-sur-Yvette, France}


\maketitle\abstracts{
The Intergalactic Magnetic Field (IGMF), permeating cosmic voids, is thought to be a relic of primordial magnetic fields generated in the early Universe and that gave rise to all astrophysical magnetic fields. While it has escaped direct detection, lower limits on its intensity can be derived by characterising the time-delayed secondary emission initiated when primary very high-energy (VHE) photons from gamma-ray bursts (GRBs) produce lepton pairs that are deflected by the IGMF before generating a secondary gamma-ray flux. Most current studies exclude IGMF values below $10^{-18}\;\mathrm{G}$, however, they are typically performed under idealised conditions. Focusing on the impact of modelling and observational choices, we simulate CTAO observations of GRBs 190114C and 221009A under varying conditions. For GRB 190114C-like sources, we establish a stable lower limit of $2\times10^{-16}\;\mathrm{G}$, robust against most variations in source properties and detection strategies. For more extreme GRB 221009A-like events, we demonstrate that CTAO could probe fields up to at least $10^{-16}\;\mathrm{G}$ under harsh conditions, improving significantly the current IGMF constraints.}

\section{Introduction}

The intergalactic magnetic field (IGMF) is the hypothetical  magnetic field that permeates cosmic voids. It is thought to be either a relic of primordial magnetic fields formed in the early Universe or a product of astrophysical phenomena \cite{Durrer_2013}. Its strength $B$ and correlation length $\lambda_B$ carry information on physical processes in the early Universe, such as inflation and phase transitions. Despite extensive efforts, clear observational evidence for the IGMF has not yet been obtained. Present lower limits on $B$ rely on its indirect effects on very-high energy (VHE, $E\gtrsim 100\;\mathrm{GeV}$) photons. $\mathrm{TeV}$ gamma rays emitted by distant sources interact with the extragalactic background light (EBL), producing electron-positron pairs that subsequently upscatter cosmic microwave background (CMB) photons up to VHE. The properties of the resulting electromagnetic cascade, or secondary flux (as opposed to the primary photons directly emitted by the source), are affected by the IGMF. It deflects the charged particles, inducing angular broadening and time delays in the secondary emission. For transient sources such as gamma-ray bursts (GRBs), the time-delay signature provides a particularly robust observable, as it does not require assumptions on long emission timescales or jet opening angles. The detection by VHE gamma-ray detectors of several GRBs at energies above a few hundred $\mathrm{GeV}$ has opened new prospects for IGMF studies. In particular, lower limits up to $10^{-18}\;\mathrm{G}$ have been established using data from GRBs 190114C and 221009A \cite{Dzhatdoev_2023}. Other works have explored the expected IGMF constraints that could be derived with the upcoming Cherenkov Telescope Array Observatory (CTAO), often under idealised assumptions \cite{Miceli_2024}. In practice, however, realistic observation conditions and specific model hypotheses introduce a number of uncertainties that directly impact the detectability of the cascade component. The purpose of our study is to assess the role of observational constraints and modelling assumptions in GRB-based IGMF studies. We focus on GRBs with properties similar to those of 190114C and 221009A, and we discuss how these underlying choices propagate into the final IGMF constraints.

\section{Method}

We simulate particle propagation using the 3D Monte Carlo code \verb|CascadEl| \footnote{\url{https://gitlab.com/rbelmont/cascapy}}, accounting for pair production and Compton scattering on the CMB and EBL within a $\Lambda\mathrm{CDM}$ cosmology. realisations. This code takes into account the temporal evolution of the intrinsic emission of the source, including the secondary photons emitted by late primary photons. The primary intrinsic source flux is modelled as a power-law in energy and time, with an unknown cut-off energy $E_\mathrm{cut}$:
\begin{equation}
\label{model}
    \Phi(E,t>t_\mathrm{min}) = \Phi_0  E^{-\gamma}t^{-\alpha}\exp(-E/E_\mathrm{cut})\;.
\end{equation}
The CTAO response is simulated using \verb|Gammapy| \cite{Donath_2023} and the official Instrument Response Functions (IRFs) \footnote{prod5 version v0.1, \url{https://doi.org/10.5281/zenodo.5499840}} for the Alpha configuration. This configuration is composed of $4$ Large Size Telescopes (LSTs) and $9$ Medium Size Telescopes (MSTs) in the North at La Palma, and $14$ MSTs and $37$ Small Size Telescopes (SSTs) in the South at Paranal. LSTs provide a $30\mathrm{s}$ slewing time and a threshold at $30\,\mathrm{GeV}$ at high altitude, and $110\;\mathrm{GeV}$ at low altitude. In comparison, MSTs slew in $90\mathrm{s}$ and have an energy threshold from $60$ to $350\;\mathrm{GeV}$, while SSTs are mainly relevant for the detection of $\mathrm{TeV}$ photons. Following the methodology of Keita et al. 2025 \cite{Keita_2025}, we generate synthetic observations across a grid of injected $B$ values and apply a time-resolved spectral fit to assess the ability of CTAO to recover them. We fit simultaneously $B$ and the intrinsic parameters ($\gamma, \alpha, E_\mathrm{cut}, \Phi_0$). We fix $\lambda_B$ to $1\;\mathrm{Mpc}$.

\section{Results}

By simulating the response of the full CTAO array to GRBs 190114C and 221009A, we evaluate the performance of our fitting procedure under both ideal and sub-optimal model assumptions and observation conditions.

\subsection{Test case: GRB 190114C}

\begin{figure*}[h]
    \centering
    \includegraphics[width=1.\linewidth]{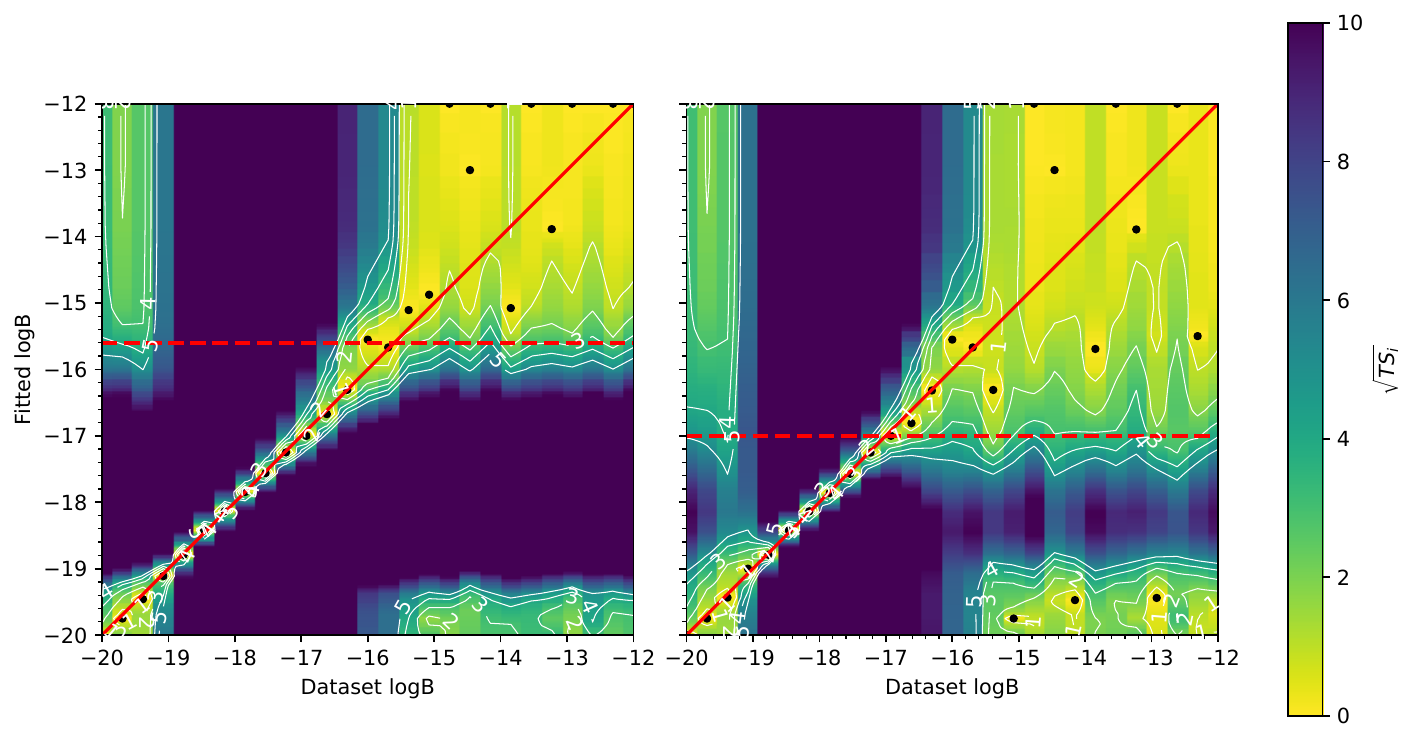}
        \caption{Confidence maps for GRB 190114C. Simulated $B$ values are shown on the abscissas and fitted values on the ordinates. Best-fit values are black dots; the colour map represents uncertainty levels up to $10\sigma$. The dashed red line highlights the $3\sigma$ lower limit for the strong field regime. Left: $E_{\mathrm{cut}}$ fit down to $10\;\mathrm{TeV}$ at least. Right: $E_{\mathrm{cut}}$ free.}
    \label{FitMap_GRB19}
\end{figure*}

GRB 190114C \cite{magic_2019}, detected at redshift $z=0.425$, serves as our primary benchmark for establishing the performance of the full array. We model its intrinsic emission with Eq. \ref{model}, using the indices measured by MAGIC ($\gamma=2.22, \alpha=1.60$) and assuming an intrinsic cut-off energy at $10\;\mathrm{TeV}$. We first assume that the VHE afterglow onset occurs at the same time as the high-energy afterglow onset detected by \textit{Fermi}-LAT in the $100\;\mathrm{MeV} - 100\;\mathrm{GeV}$ energy band, at $t_\mathrm{min}=6\mathrm{s}$ after the burst time $T_0$. We still add a $60\mathrm{s}$ delay before observation, to account for the maximum slewing time of LSTs. In the strong IGMF regime where the cascade is not detectable, there is a degeneracy between $E_{cut}$ and $B$ as both can lead to a deficit of VHE photons. It can happen either through an intrinsic lack of VHE emission (small $E_\mathrm{cut}$) or through a too strong cascade dilution (large $B$). Based on recent LHAASO observations \cite{cao_2023} of photons up to $13\;\mathrm{TeV}$ for GRB 221009A, it is physically reasonable to assume $E_{cut} \geq 10\;\mathrm{TeV}$. Consequently, we apply this prior to our test case, initially forbidding the fit from searching for a cut-off energy below $10\;\mathrm{TeV}$. The left picture of Fig.~\ref{FitMap_GRB19} shows with black dots the best fitted $\log B$ (ordinate) for every dataset with injected $\log B$ in the abscissa. The fit procedure encounters three distinct regimes. In the intermediate regime ($10^{-19}\;\mathrm{G} < B < 10^{-16} \;\mathrm{G}$), the cascade component falls within the CTAO sensitivity window, allowing for robust constraints on $B$. Outside this optimal window, the cascade signal is essentially lost. For very weak fields ($B < 10^{-19}\;\mathrm{G}$), the cascade is hidden because its peak energy moves rapidly below the CTAO energy threshold. Conversely, for very strong fields ($B > 10^{-16}\;\mathrm{G}$), the secondary emission is too suppressed to be distinguishable from the much stronger primary flux.

\begin{table*}
    \centering
    \begin{tabular}{l l}
        \hline\hline
        \textbf{Study} & \textbf{Lower Limit ($\mathrm{G}$)}\\
        \hline
        Reference & $2 \times 10^{-16}$ \\
        Reduced fluence: $\phi_0/5$& $2 \times 10^{-16}$\\
        \textbf{Reduced fluence:  $\phi_0/10$}& $10^{-17}$ \\
        \textbf{Reduced fluence:  $< \phi_0/10$}& no constraint\\
        $\alpha$ and $\gamma$ varying within published uncertainties& $2 \times 10^{-16}$\\
        \textbf{$\phi_0/5$ and strong variations in $\alpha$ and $\gamma$}& $3 \times 10^{-16}$\\
        \textbf{$E_\mathrm{cut}$ let free in the fit}& $10^{-17}$ \\
        Detection delay increased up to $1\,\mathrm{hr}$ & $2 \times 10^{-16}$\\
        \textbf{Afterglow signal starts $1'$ later}& $4 \times 10^{-17}$ \\
        High zenith angle (lower energy thresholds) & $2 \times 10^{-16}$\\ 
        Moonlight data taking (increased energy thresholds) & $2 \times 10^{-16}$\\
        \textbf{Adding LSTs in the South}& $5 \times 10^{-16}$ \\
        Exchanging North and South arrays & $2 \times 10^{-16}$\\
        Change in the fit model & $2 \times 10^{-16}$\\
        Omitting South site& $2\times10^{-16}$ \\
        \textbf{Omitting North site}& $10^{-17}$ \\
    \end{tabular}
    \caption{Varying assumptions for GRB 190114C. Bold cases indicate variations that impact the lower limit.}
    \label{tab:systematics}
\end{table*}

We then performed a comprehensive study varying intrinsic properties and observational strategies, summarised in Table~\ref{tab:systematics}. Two specific assumptions are particularly liable to degrade the results. Firstly, if $E_{\mathrm{cut}}$ is left as a free parameter in the fit rather than being constrained to $E_{\mathrm{cut}} \geq 10\;\mathrm{TeV}$, the lower limit is degraded to $10^{-17}\;\mathrm{G}$, as illustrated in the right panel of Fig.~\ref{FitMap_GRB19}. This occurs because the likelihood fit may interpret the absence of a low-energy cascade as an intrinsic lack of VHE primary photons and not necessarily the presence of a particularly strong IGMF. Secondly, the onset time of the VHE afterglow dictates the total energy available to initiate the cosmological cascade via the primary fluence. Shifting the assumed VHE afterglow onset from $T_0 + 6\mathrm{s}$ to $T_0 + 62\mathrm{s}$ (the start of MAGIC observations in the $100\;\mathrm{GeV}- 10\;\mathrm{TeV}$ band) reduces the total available fluence by a factor of $4.5$. This directly weakens the secondary emission signal, leading to a reduced lower limit of $4 \times 10^{-17}\;\mathrm{G}$.

\subsection{Extreme case: GRB 221009A}

GRB 221009A ($z=0.151$) represents an extreme case \cite{cao_2023}. Following LHAASO observations, the intrinsic flux is modelled as a power law in energy with a temporal evolution consisting of four consecutive power-law segments, normalised to LHAASO measurements. Similarly to GRB 190114C, a $10\;\mathrm{TeV}$ cut-off energy is assumed, as it is compatible with the data. In ideal conditions without Moonlight, CTAO is expected to enable the probing of IGMF strengths up to $10^{-15}\;\mathrm{G}$. However, the presence of strong Moonlight during the actual event raises the energy threshold to $100\;\mathrm{GeV}$ in the North. This degrades the lower limit to $\approx 3 \times 10^{-16}\;\mathrm{G}$. Furthermore, if the first night of observation is missed, the secondary emission for $B > 10^{-16}\;\mathrm{G}$ becomes indistinguishable from the background, resulting in a total degeneracy between $B$ and $E_{\mathrm{cut}}$. While LST-1 observations already suggest a lower limit of $B \gtrsim 3 \times 10^{-17}\;\mathrm{G}$, the full CTAO array would enhance this by an order of magnitude, even under these harsh conditions. This would significantly improve upon the current $10^{-17}\;\mathrm{G}$ limits obtained with \textit{Fermi}-LAT \cite{Dzhatdoev_2023}.

\section{Conclusion}

For GRBs like 190114C, CTAO can probe magnetic field strengths up to $2 \times 10^{-16}\;\mathrm{G}$, with constraints remaining stable across a wide range of observational uncertainties. In the case of exceptional events like GRB 221009A, CTAO sensitivity extends even further, potentially reaching $10^{-15}\;\mathrm{G}$ under optimal conditions and $10^{-16}\;\mathrm{G}$ if there is strong Moonlight. In any case, our study demonstrates that CTAO is guaranteed to increase IGMF constraints up to $4\times 10^{-17}\;\mathrm{G}$ at least, assuming $E_\mathrm{cut}\geq 10\;\mathrm{TeV}$ and $\lambda_B=1\;\mathrm{Mpc}$.

\section*{Acknowledgment}

\noindent We gratefully acknowledge financial support from the agencies and organizations listed at \href{https://www.ctao.org/for-scientists/library/acknowledgments/}{https://www.ctao.org/for-scientists/library/acknowledgments/}. This research used the CTAO IRFs provided by the CTAO Consortium and Observatory, see \href{https://www.cta-observatory.org/science/ctao-performance}{https://www.cta-observatory.org/science/ctao-performance}.

\section*{References}
\bibliography{keita}

@inproceedings{Keita_2025,
   title={Expected Constraints on the Intergalactic Magnetic Field using Gamma-Ray Bursts with the Cherenkov Telescope Array Observatory},
   journal = "PoS",
   author={Keita, T. and others},
   year={2025},
   pages={710}}

@article{magic_2019,
   title={Teraelectronvolt emission from the gamma-ray burst GRB 190114C},
   volume={575},
   number={7783},
   journal={Nature},
   publisher={Springer Science and Business Media LLC},
   author={Acciari, V. A. and others},
   year={2019},
   month=nov, pages={455–458} }

@article{cao_2023,
   title={Very high-energy gamma-ray emission beyond 10 TeV from GRB 221009A},
   volume={9},
   number={46},
   journal={Science Advances},
   publisher={American Association for the Advancement of Science (AAAS)},
   author={Cao, Z. and others},
   year={2023},
   month=nov }

@article{Donath_2023,
   title={Gammapy: A Python package for gamma-ray astronomy},
   volume={678},
   journal={Astronomy \& Astrophysics},
   publisher={EDP Sciences},
   author={Donath, A. and others},
   year={2023},
   month=oct, pages={A157} }

@article{Dzhatdoev_2023,
   title={First constraints on the strength of the extragalactic magnetic field from gamma-ray observations of GRB 221009A},
   volume={527},
   journal={Monthly Notices of the Royal Astronomical Society: Letters},
   publisher={Oxford University Press (OUP)},
   author={Dzhatdoev, T. A. and others},
   year={2023},
   month=oct, 
   pages={L95–L102} }

@article{Miceli_2024,
   title={Prospects for detection of the pair-echo emission from TeV gamma-ray bursts},
   volume={688},
   ISSN={1432-0746},
   url={http://dx.doi.org/10.1051/0004-6361/202449305},
   DOI={10.1051/0004-6361/202449305},
   journal={Astronomy and Astrophysics},
   publisher={EDP Sciences},
   author={Miceli, D. and Da Vela, P. and Prandini, E.},
   year={2024},
   month=aug, pages={A57} }

@article{Durrer_2013,
   title={Cosmological magnetic fields: their generation, evolution and observation},
   volume={21},
   number={1},
   journal={The Astronomy and Astrophysics Review},
   publisher={Springer Science and Business Media LLC},
   author={Durrer, Ruth and Neronov, Andrii},
   year={2013},
   month={jun} }


\end{document}